\documentclass[11pt,a4paper]{article}

\usepackage[margin=1in]{geometry}
\usepackage{amsmath,amssymb}
\usepackage{graphicx}
\usepackage{booktabs}
\usepackage{hyperref}
\usepackage{siunitx}
\usepackage{subcaption}

\hypersetup{
    colorlinks=true,
    linkcolor=blue,
    citecolor=blue,
    urlcolor=blue,
}

\title{Physics-Informed Neural Networks for Static Black-Hole Exterior Metrics:\\
Charge and Cosmological-Constant Sweeps}

\author{Huan Jin$^{a}$\\[4pt]
Fei Wu\\
\href{mailto:f.eiwu@outlook.com}{f.eiwu@outlook.com}\\[4pt]
Fei Xue$^{a}$\\
\href{mailto:xuefei_work@126.com}{xuefei\_work@126.com}\\[8pt]
{\small $^{a}$School of Information Engineering, Jiangxi Institute of Technology,
Nanchang, Jiangxi 330098, PR China}}

\date{\today}

\begin{document}

\maketitle

\begin{abstract}
We apply physics-informed neural networks (PINNs) to recover the time-time component
of static, spherically symmetric black-hole exterior metrics from a reduced ordinary
differential equation (ODE). Inspired by recent work on solving Einstein field
equations with deep learning~\cite{Li2023}, we encode the vacuum/charged exterior
through a residual loss and an asymptotic boundary constraint---not by embedding
analytic metric terms such as $2M/r$ directly into the network output. Unlike the distributed
PINN (DPINN) strategy of Ref.~\cite{Li2023}, which partitions the radial domain into
subdomains with separate networks, we employ a \emph{unified} fully connected network
over the entire interval $[r_{\min},r_{\max}]$, avoiding spurious jumps at subdomain
interfaces. Moreover, whereas Ref.~\cite{Li2023} restricts training to
$r\in(10,300)$---well outside the steep $1/r$ and $1/r^2$ curvature of the inner
exterior---we begin at $r_{\min}=10^{-2}M$, spanning three decades in radius
and covering the strongly varying region that DPINNs sidestep by domain truncation.
Holding the mass fixed at
$M=1$, we sweep electric charge $Q\in\{0,0.5,1.0,1.1\}$ at $\Lambda=0$ and sweep
$\Lambda\in\{0,0.1,-0.1\}$ at $Q=0.5$. All configurations achieve relative
$\mathcal{L}_2$ errors below $6\%$ against the analytic reference. For the
representative case $Q=0.5$, $\Lambda=0.1$, three independent trainings with fixed
seeds yield relative errors of $3.89\%$, $1.08\%$, and $2.76\%$ (mean $2.58\%$,
standard deviation $1.41\%$), demonstrating robustness of the mesh-free approach
without labeled field data.
\end{abstract}

\section{Introduction}
\label{sec:intro}

Einstein field equations couple spacetime geometry to matter through a nonlinear
tensor system~\cite{Wald1984,Carroll2004,MTW1973}. Exact solutions---from
Schwarzschild and Kerr black holes to charged and cosmological extensions---anchor
theoretical astrophysics~\cite{Schwarzschild1916,Kerr1963,Reissner1918,deSitter1917,Chandrasekhar1983,HawkingEllis1973,Poisson2004} but are available only in
special symmetry classes. Astrophysical inference from gravitational-wave
detections~\cite{Abbott2016,Abbott2017} and horizon-scale imaging~\cite{EHT2019,EHT2022}
motivates flexible forward solvers that can explore parameter dependence beyond
tabulated metrics. Numerical relativity traditionally relies on $3+1$ decompositions,
adaptive mesh refinement, and careful treatment of gauge and boundary
conditions~\cite{Pretorius2005,Baumgarte2010,Shibata2009,Alcubierre2008,Lehner2001,Gundlach2007,Gourgoulhon2012,Font2000}.
Physics-informed neural networks (PINNs)~\cite{Raissi2019,Karniadakis2021,Lu2021,Cuomo2022} and related
scientific machine-learning frameworks~\cite{E2018,Sirignano2018,Liao2019,Raissi2020,Haghighat2021,Wang2021,Cai2021}
offer an alternative: a neural network represents the field, PDE residuals and boundary
conditions enter the loss, and derivatives are obtained via automatic differentiation.
Domain-decomposition variants (XPINNs, cPINNs)~\cite{Jagtap2020,Shukla2021,Jagtap2020cPINN}
and operator-learning approaches~\cite{Li2020FNO,Kovachki2021,Lu2021DeepONet}
extend this paradigm, though training pathologies and stiffness remain active
research topics~\cite{Krishnapriyan2021,Wang2022,McClenny2020,Wight2020,Song2021}.

Machine learning has recently been applied to Einstein's equations, gravitational
waves, and black-hole perturbation theory~\cite{Li2023,Cornell2022,Luna2022,Patel2024,Kohli2020,Schmidt2021,Lee2021,Bhagwat2021,George2017,Berti2009,Teukolsky1973}.
Most prior work either reconstructs multiple metric components from Ricci residuals or
targets specialized wave equations; reduced radial models remain useful benchmarks for
architecture and domain-design choices before scaling to full tensor systems.

Li \textit{et al.}~\cite{Li2023} demonstrated that PINNs can recover Schwarzschild
and charged Schwarzschild metrics by minimizing Ricci-tensor residuals. To improve
convergence on stiff radial profiles, they adopt a distributed PINN (DPINN)
formulation: the domain is split into overlapping or adjacent subintervals, each
fitted by an independent network. As shown in their charged-Schwarzschild error
curves (reproduced in Fig.~\ref{fig:li2023}), this produces visible discontinuities
in the pointwise $\mathcal{L}_2$ error at subdomain boundaries---typically where
separate optimizers meet. A further simplification in that reference is the radial
cutoff: their Schwarzschild and charged-Schwarzschild experiments integrate the
physics loss only on $r\in(10,300)$, dividing this interval into three or four
subdomains with a separate network each. This excludes the inner domain where
Eq.~\eqref{eq:reference} varies most rapidly ($A\sim 1-2M/r+\cdots$ as $r\to 0$).
They also build the metric components from hand-crafted analytic templates: for
Schwarzschild, Ref.~\cite{Li2023} set
$g_{00}=u_0(r)/r^2+2M/r-1$ and $g_{11}=u_1(r)$, so the leading asymptotic behavior
$g_{00}\to-(1-2M/r)$ as $r\to\infty$ is injected into the architecture rather than
learned from the field equations and a boundary penalty alone. Here we pursue a
complementary reduced formulation: a single radial ODE for the metric function
associated with $g_{tt}$, extended to nonzero cosmological constant $\Lambda$, with
the minimal ansatz $A(r)=u(r)/r^2$ and asymptotic information entering only through
the soft constraint~\eqref{eq:bc} at $r_{\max}$. The same network template therefore
applies across our charge and $\Lambda$ sweeps without redesigning the output map
when the analytic metric changes. The problem is solved by one unified network from
$r_{\min}=10^{-2}M$ outward. The resulting problem is substantially stiffer---the
network must simultaneously fit steep inner gradients and asymptotic outer
behavior---but more representative of a global exterior solve and less reliant on
prior knowledge of the metric's functional form. We focus on parameter sweeps over
charge and $\Lambda$, convergence diagnostics from TensorBoard, and robustness
across random seeds. Our goal is not a full $g_{\mu\nu}$ reconstruction but a
controlled benchmark showing how a unified PINN with a minimal ansatz delivers
smooth metric profiles without interface artifacts on a domain that includes the
curvature-dominated inner region.

\section{Method}
\label{sec:method}

\subsection{Metric ansatz and reference solution}

For a static, spherically symmetric line element we write
\begin{equation}
    ds^2 = -f(r)\,dt^2 + g(r)\,dr^2 + r^2(d\theta^2 + \sin^2\theta\,d\varphi^2),
    \label{eq:metric}
\end{equation}
and identify the time-time component through $f(r) = -g_{tt}(r)$.
In the present PINN formulation, we introduce a scalar network
output $u(r)$ and relate it to the metric function
\begin{equation}
    A(r) \equiv -g_{tt}(r) = \frac{u(r)}{r^2}.
    \label{eq:ansatz}
\end{equation}
The analytic reference used for evaluation is
\begin{equation}
    A_{\mathrm{ref}}(r) = 1 - \frac{2M}{r} + \frac{Q^2}{r^2} - \frac{\Lambda r^2}{3},
    \label{eq:reference}
\end{equation}
which interpolates Schwarzschild ($Q=0$, $\Lambda=0$)~\cite{Schwarzschild1916},
Reissner--Nordstr\"om ($Q\neq 0$, $\Lambda=0$)~\cite{Reissner1918}, and a de~Sitter-like
correction at large $r$ when $\Lambda\neq 0$~\cite{deSitter1917}. Natural units with $G=c=1$ are implied; we set $M=1$ throughout.

\subsection{Minimal ansatz versus hard-coded metric form}
\label{sec:ansatz}

Ref.~\cite{Li2023} construct the metric tensor from network outputs using
problem-specific analytic additions. In their Schwarzschild example,
\begin{equation}
    g_{00} = \frac{u_0(r)}{r^2} + \frac{2M}{r} - 1,
    \qquad g_{11} = u_1(r),
    \label{eq:li-ansatz}
\end{equation}
so the dominant asymptotic piece $2M/r-1$ is inserted by hand; the network learns
only a correction $u_0(r)/r^2$ around a metric form that already resembles the
expected solution. Similar charged-metric templates are used for
$g_{00}$ in their Reissner--Nordstr\"om runs. In effect, much of the boundary
information is encoded in the \emph{architecture}, reducing what the loss must infer
from the Einstein residuals alone.

Our approach keeps the output map generic. Equation~\eqref{eq:ansatz} uses only the
spherical symmetry factor $1/r^2$ relating $u(r)$ to $A(r)=-g_{tt}(r)$; no
$2M/r$, $Q^2/r^2$, or $\Lambda r^2$ terms are added to the network output. All
asymptotic information enters through the soft penalty~\eqref{eq:bc}, which asks
that $A(r_{\max})$ match the leading large-$r$ behavior implied by
Eq.~\eqref{eq:reference}. The interior profile is controlled entirely by the ODE
residual~\eqref{eq:residual} with parameters $(M,Q,\Lambda)$ supplied to the loss,
not hard-wired into $u_\theta$. This separation has two practical advantages.
First, the same network architecture applies to every entry in our charge and
$\Lambda$ sweeps without modifying the output ansatz when the analytic metric
changes. Second, the trained $u_\theta$ must genuinely satisfy the physics over the
full domain---including the inner region---rather than inheriting the outer
Schwarzschild tail from Eq.~\eqref{eq:li-ansatz}.

\subsection{ODE residual and boundary condition}

The network is trained without labeled grids of $A(r)$. Instead, automatic
differentiation enforces a radial residual (as implemented in the loss module):
\begin{equation}
    \mathcal{R}(r) = r\,\frac{du}{dr} - u - r^2 + Q^2 + \Lambda r^4 = 0.
    \label{eq:residual}
\end{equation}
The mean-squared residual $\mathbb{E}_r[\mathcal{R}^2]$ forms the physics term in
the loss. At the outer radius $r_{\max}$ of the computational domain we impose an
asymptotic constraint consistent with Eq.~\eqref{eq:reference}:
\begin{equation}
    \mathcal{B} = \frac{u(r_{\max}) - r_{\max}^2 + \Lambda r_{\max}^4/3}
    {2 M r_{\max}} + 1,
    \label{eq:bc}
\end{equation}
with penalty $10\,\mathcal{B}^2$ added to the residual loss. Collocation points are
drawn on a log-spaced radial grid from $r_{\min}=10^{-2}M$ to
$r_{\max}=\max(10M,\,15Q^2/(2M))$. Table~\ref{tab:domain} contrasts this domain
with Ref.~\cite{Li2023}. At $M=1$ our inner radius lies more than three orders of
magnitude below theirs; the reference metric and its derivatives are therefore
much steeper near $r_{\min}$ than at $r=10$. Mapping the input through $\log r$
mitigates but does not remove this stiffness: the unified network must represent
both the inner $1/r$-dominated regime and the outer asymptotically flat (or
$\Lambda$-modified) region with a single weight vector.

\begin{table}[htbp]
    \centering
    \caption{Radial domains: Ref.~\cite{Li2023} versus this work ($M=1$).}
    \label{tab:domain}
    \begin{tabular}{@{}lcc@{}}
        \toprule
        & Li \textit{et al.}~\cite{Li2023} & This work \\
        \midrule
        Inner radius $r_{\min}$ & $10$ & $10^{-2}$ \\
        Outer radius $r_{\max}$ & $300$ & $\max(10,\,15Q^2/2)$ \\
        Subdomains & 3 (Schw.) / 4 (ch.) & 1 (unified) \\
        Inner curvature regime & excluded & included \\
        Metric output map & $g_{00}=u_0/r^2+2M/r-1$ & $A=u/r^2$ only \\
        Asymptotic BC & hard-coded in ansatz & soft loss at $r_{\max}$ \\
        \bottomrule
    \end{tabular}
\end{table}

\subsection{Unified vs.\ distributed PINN}
\label{sec:unified}

\begin{figure}[htbp]
    \centering
    \includegraphics[width=0.95\linewidth]{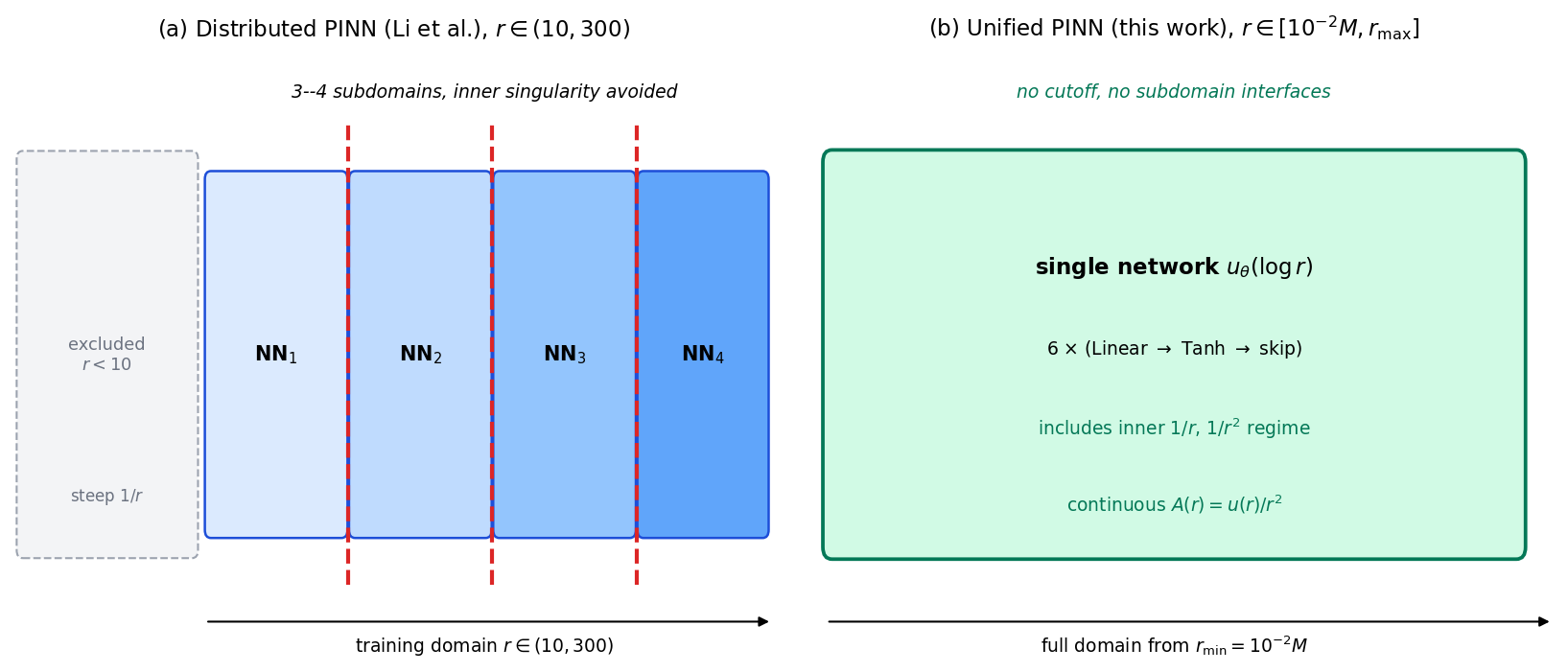}
    \caption{Architectural comparison. (a)~Distributed PINN~\cite{Li2023} on
    $r\in(10,300)$: independent networks on adjacent subdomains; the inner region
    $r<10$ (shaded) is excluded from training. (b)~Unified PINN (this work) on
    $[10^{-2}M,r_{\max}]$: one network over the full domain including the steep
    inner curvature regime, yielding a continuous $A(r)$.}
    \label{fig:arch-compare}
\end{figure}

Figure~\ref{fig:arch-compare} contrasts the two architectural philosophies.
In the DPINN approach of Ref.~\cite{Li2023}, the radial line is decomposed into
several patches on $r\in(10,300)$, each with its own network and loss minimization.
The inner region $r<10$, where the metric coefficients vary most sharply, is not
part of the training domain at all. Interface
continuity is not enforced exactly, so the reconstructed metric components can
exhibit kinks and the local error displays sharp steps (Fig.~\ref{fig:li2023},
right panel). By contrast, our unified PINN assigns a \emph{single} set of weights
to represent $u_\theta(\log r)$ on $[10^{-2}M,\,r_{\max}]$, including the
inner exterior where $2M/r$ and $Q^2/r^2$ dominate.
Collocation points are drawn from one log-spaced sampler; residual and boundary
losses are evaluated on the same network everywhere. Because $u_\theta$ is a
continuous function of $r$, the derived metric function $A(r)=u/r^2$ is smooth
across the exterior without hand-crafted gluing between subdomains.

\begin{figure}[htbp]
    \centering
    \includegraphics[width=0.98\linewidth]{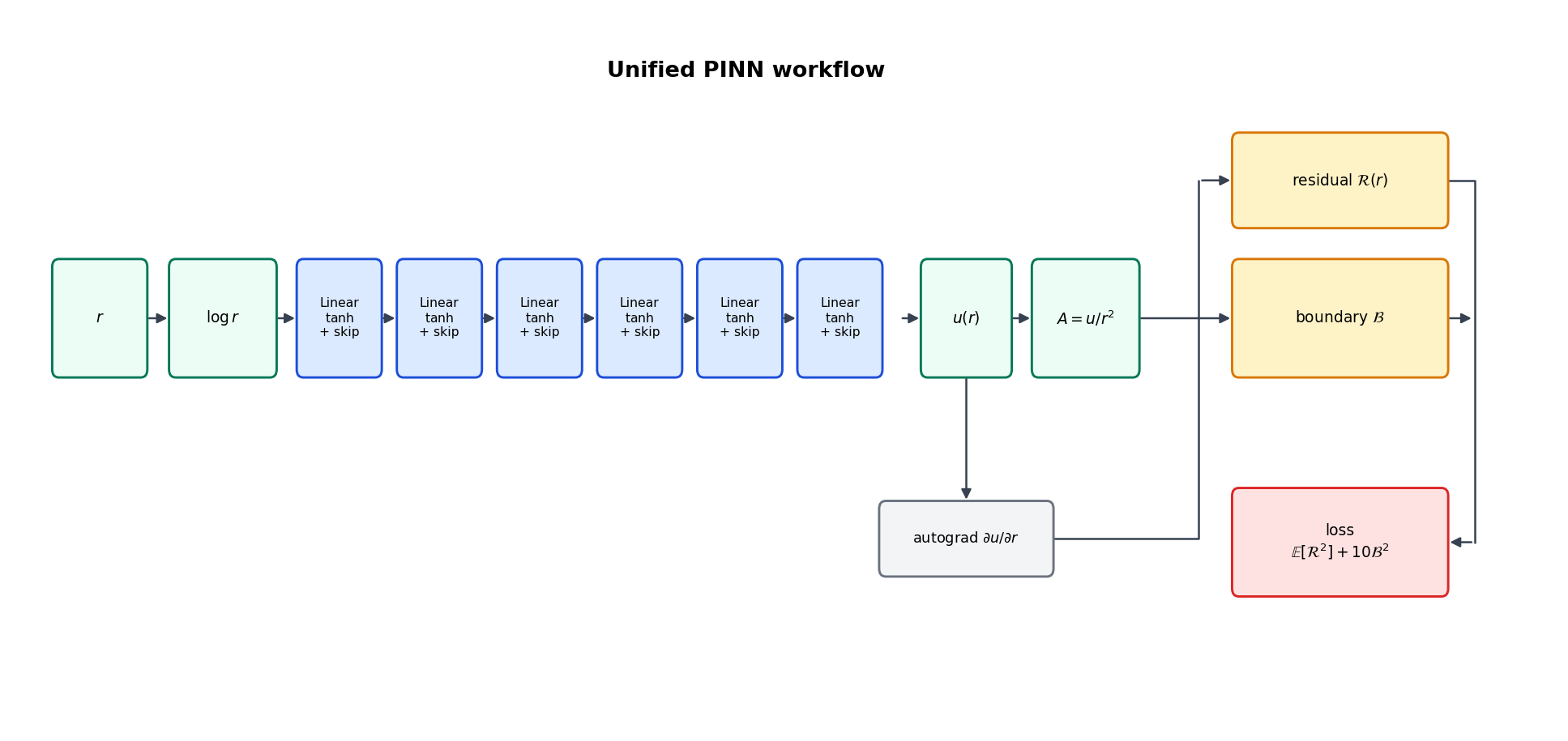}
    \caption{Unified PINN workflow. A single network maps $\log r$ to $u(r)$ with the
    generic ansatz $A(r)=u(r)/r^2$; asymptotic behavior enters only through the
    boundary penalty~\eqref{eq:bc}, not through hand-added metric terms. Six
    width-4 blocks with additive skip connections---included to improve training
    convergence---are used throughout.}
    \label{fig:unified-arch}
\end{figure}

Figure~\ref{fig:unified-arch} details the unified workflow. The scalar input $\log r$ passes through six residual blocks
(each consisting of a linear map, $\tanh$ activation, and additive skip from the
previous layer), producing $u(r)$. The skips are included specifically to aid
optimization convergence on this stiff, log-spaced domain: they provide shortcut
gradient paths through the deep $\tanh$ stack and mitigate signal attenuation when
fitting both the steep inner profile and the outer asymptotic behavior in a single
network~\cite{He2016}. Automatic differentiation supplies
$\mathrm{d}u/\mathrm{d}r$ for the ODE residual~\eqref{eq:residual} and for the
asymptotic boundary penalty~\eqref{eq:bc}. All physics terms back-propagate through
the same computational graph, so no subdomain boundary conditions are required.

\subsection{Network architecture and optimization}

\begin{figure}[htbp]
    \centering
    \includegraphics[width=0.98\linewidth]{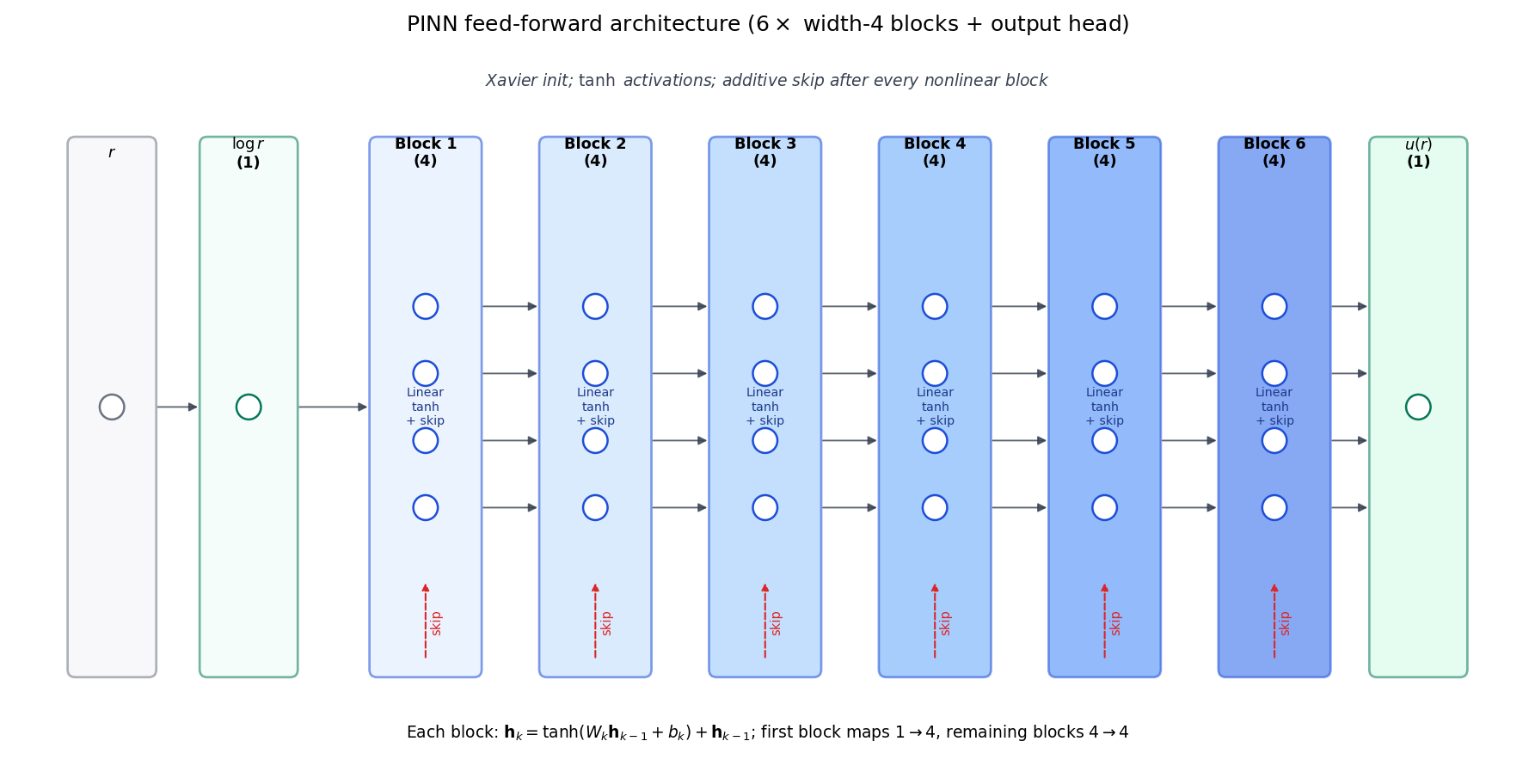}
    \caption{Feed-forward PINN architecture: input $r$, transform to $\log r$, six
    width-4 blocks with $\tanh$ and additive skip connections (included to aid
    convergence), and scalar output $u(r)$. Dashed red arrows indicate the skip
    path inside each block.}
    \label{fig:network-layers}
\end{figure}

The PINN maps $\log r$ through six hidden layers of width four with $\tanh$
activations and additive skip connections between nonlinear blocks. Unlike
Ref.~\cite{Li2023}, who improve convergence mainly by partitioning the radial
domain into separate networks, we retain a single unified approximator and instead
add skips after every nonlinear block to ease optimization over the full interval
$[r_{\min},r_{\max}]$. Each block has the form
$\mathbf{h}_k=\tanh(W_k\mathbf{h}_{k-1}+b_k)+\mathbf{h}_{k-1}$ (with the first
linear map $1\to4$ and subsequent maps $4\to4$), so lower layers retain a direct
influence on later representations and on the scalar output $u(r)$; in practice this
stabilizes residual and boundary-loss descent when the ODE stiffness varies by orders
of magnitude across the collocation grid. Figure~\ref{fig:network-layers}
shows the layer structure: an input transform
$r\mapsto\log r$, six such blocks, and a final linear head $4\to1$
producing $u(r)$. Weights are initialized with Xavier uniform~\cite{Glorot2010}; biases are zero.
The additive skip structure follows residual-network practice~\cite{He2016}, where
shortcut connections are introduced primarily to aid convergence in deep networks.

\begin{figure}[htbp]
    \centering
    \includegraphics[width=0.98\linewidth]{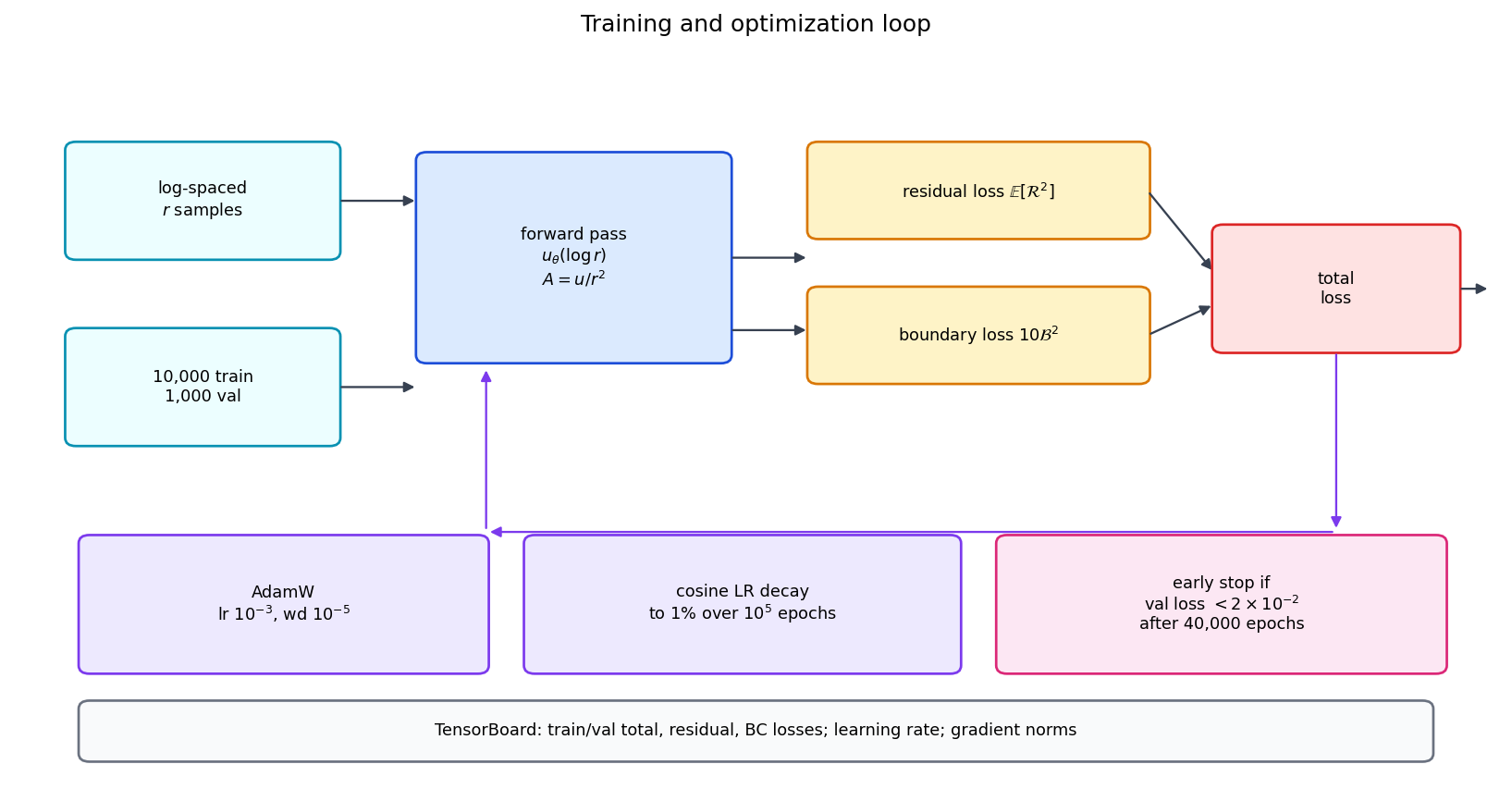}
    \caption{Training and optimization loop: log-spaced collocation, physics-informed
    loss evaluation, AdamW with cosine decay, and early stopping on validation loss.}
    \label{fig:training}
\end{figure}

Figure~\ref{fig:training} summarizes the optimization loop. Each epoch draws
$10{,}000$ training and $1{,}000$ validation collocation points on a log-spaced
grid. The total loss combines residual and boundary terms; AdamW~\cite{Kingma2015,Loshchilov2019}
($\mathrm{lr}=10^{-3}$, weight decay $10^{-5}$) updates the unified network with a
cosine learning-rate decay to $1\%$ of the initial rate over $10^5$ epochs. Training
stops early when the best validation loss falls below $2\times 10^{-2}$ after at
least $40{,}000$ epochs. TensorBoard logs record total, residual, and boundary
losses, learning rate, and gradient norms.

\subsection{Error metric}

After training, we evaluate the relative $\mathcal{L}_2$ error on $500$ log-spaced
points:
\begin{equation}
    \varepsilon_{\mathrm{rel}}
    = \frac{\|A_{\mathrm{pred}} - A_{\mathrm{ref}}\|_2}
           {\|A_{\mathrm{ref}}\|_2},
    \label{eq:l2}
\end{equation}
where $A_{\mathrm{pred}}(r)=u(r)/r^2$.

\section{Experiments}
\label{sec:experiments}

We conduct two sweeps at fixed $M=1$:

\begin{itemize}
    \item \textbf{Charge sweep} ($\Lambda=0$): $Q\in\{0,\,0.5,\,1.0,\,1.1\}$.
    \item \textbf{$\Lambda$ sweep} ($Q=0.5$): $\Lambda\in\{0,\,0.1,\,-0.1\}$.
\end{itemize}

The $\Lambda=0$, $Q=0.5$ case appears in both series. A separate \textbf{robustness}
study repeats the $Q=0.5$, $\Lambda=0.1$ configuration three times with seeds
$101$, $102$, and $103$.

\section{Results}
\label{sec:results}

\subsection{Parameter sweeps}

Table~\ref{tab:results} summarizes relative $\mathcal{L}_2$ errors, best validation
loss, and stopping epoch for each configuration. All runs satisfy
$\varepsilon_{\mathrm{rel}} < 0.06$.

\begin{table}[htbp]
    \centering
    \caption{PINN results at $M=1$. $\varepsilon_{\mathrm{rel}}$ is the relative
    $\mathcal{L}_2$ error against Eq.~\eqref{eq:reference}; val.\ loss is the best
    validation loss; epoch is the best-validation epoch.}
    \label{tab:results}
    \begin{tabular}{@{}lcccccc@{}}
        \toprule
        Tag & $Q$ & $\Lambda$ & $\varepsilon_{\mathrm{rel}}$ & Val.\ loss & Epoch \\
        \midrule
        Q0\_L0   & 0.0  & 0.0  & 0.0498 & $4.54\times 10^{-4}$ & 39\,978 \\
        Q05\_L0  & 0.5  & 0.0  & 0.0128 & $2.66\times 10^{-3}$ & 19\,973 \\
        Q1\_L0   & 1.0  & 0.0  & 0.0061 & $5.86\times 10^{-2}$ & 9\,953 \\
        Q11\_L0  & 1.1  & 0.0  & 0.0276 & $5.82\times 10^{-2}$ & 10\,279 \\
        Q05\_L01 & 0.5  & 0.1  & 0.0594 & $5.87\times 10^{-2}$ & 53\,990 \\
        Q05\_Lm01& 0.5  &$-0.1$& 0.0560 & $5.92\times 10^{-1}$ & 97\,652 \\
        \bottomrule
    \end{tabular}
\end{table}

\begin{figure}[htbp]
    \centering
    \includegraphics[width=0.92\linewidth]{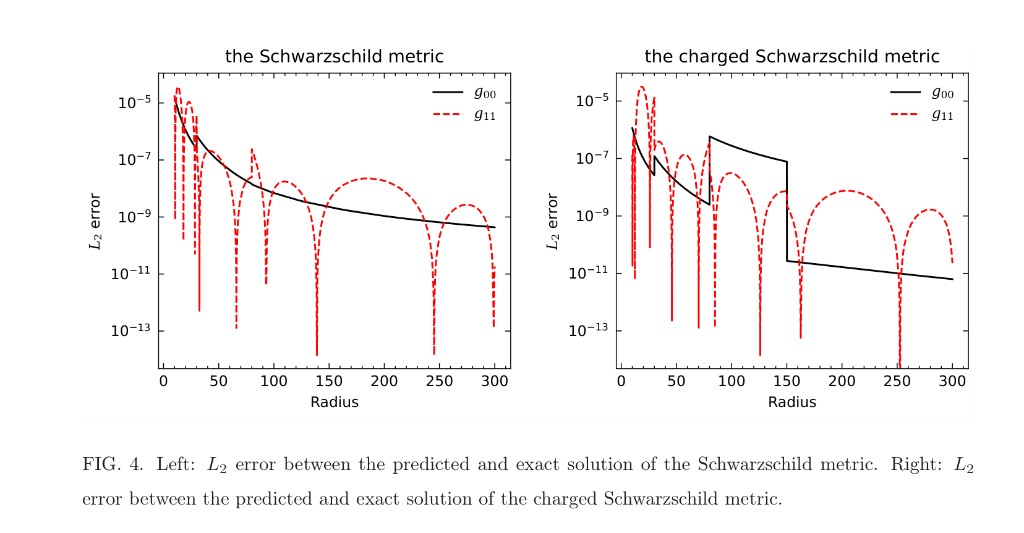}
    \caption{Reference DPINN error profiles from Li \textit{et al.}~\cite{Li2023}
    (their Fig.~4): pointwise $\mathcal{L}_2$ error versus radius for Schwarzschild
    (left) and charged Schwarzschild (right). Vertical jumps in the charged case
    mark subdomain interfaces. Our unified network avoids this artifact by
    construction.}
    \label{fig:li2023}
\end{figure}

\begin{figure}[htbp]
    \centering
    \includegraphics[width=0.95\linewidth]{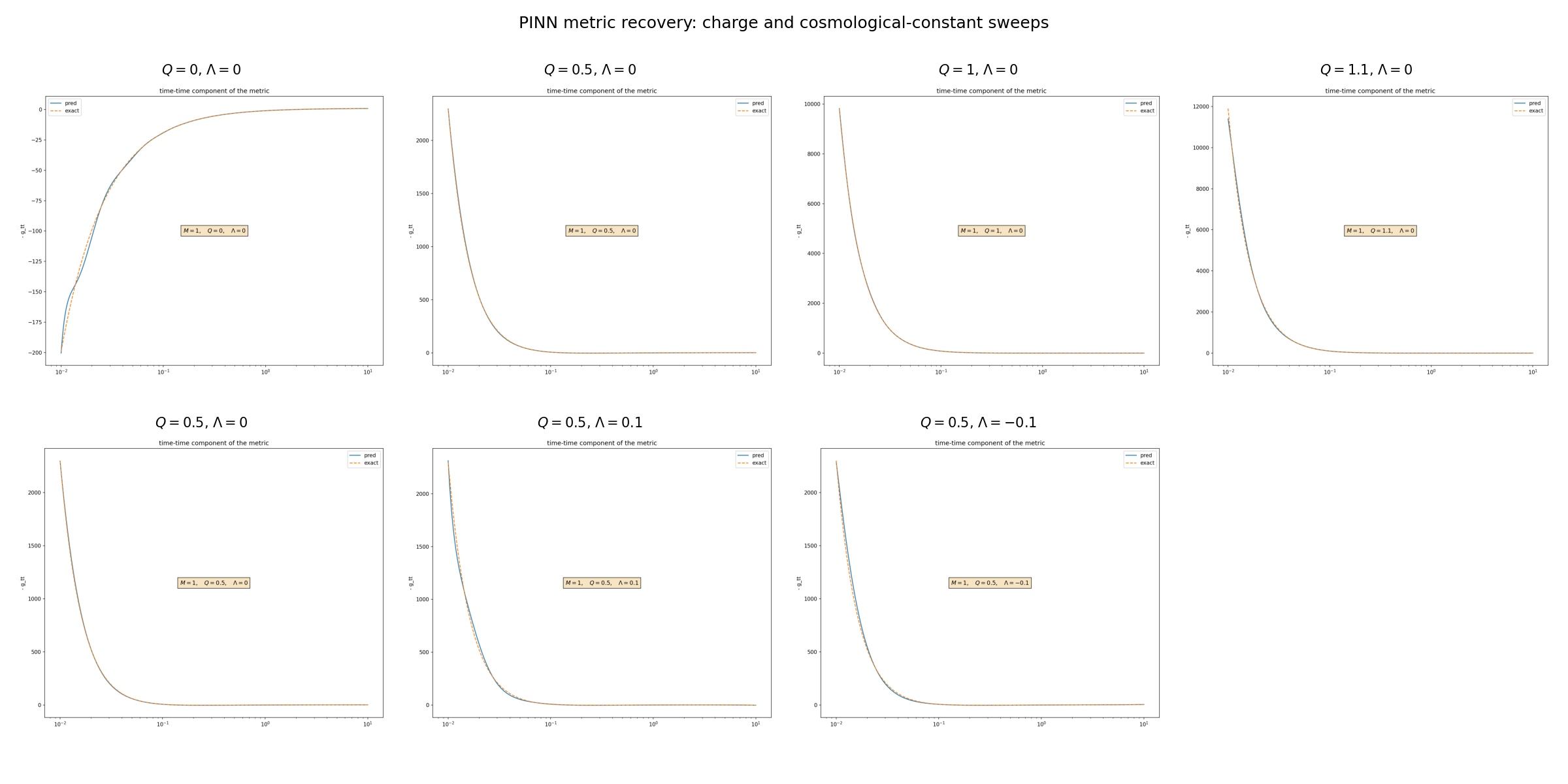}
    \caption{Unified PINN prediction (solid) versus analytic reference (dashed) for
    $-g_{tt}$ across charge and $\Lambda$ sweeps ($M=1$). Curves are continuous over
    the full domain.}
    \label{fig:comparison}
\end{figure}

Figure~\ref{fig:comparison} overlays PINN predictions and analytic references for
all six tags on the full domain down to $r_{\min}=10^{-2}M$. The predicted curves
are visually smooth on log-spaced $r$; no step discontinuities appear at interior
radii, in contrast to the DPINN error profile of Ref.~\cite{Li2023}
(Fig.~\ref{fig:li2023}), which is reported only for $r>10$. Resolving the inner
decade---where $|A_{\mathrm{ref}}|$ changes most rapidly---is the dominant
numerical difficulty in our sweeps; the unified architecture succeeds without
partitioning this region off. At $\Lambda=0$, increasing $Q$ from Schwarzschild to
near-extremal charge ($Q=1.1$) changes the exterior curvature; the unified PINN
tracks the reference without re-meshing or subdomain stitching. For $Q=0.5$,
positive $\Lambda$ steepens the large-$r$ slope of $-g_{tt}$ whereas negative
$\Lambda$ flattens it; both signs are captured with comparable accuracy.

\subsection{Training convergence}

\begin{figure}[htbp]
    \centering
    \includegraphics[width=0.95\linewidth]{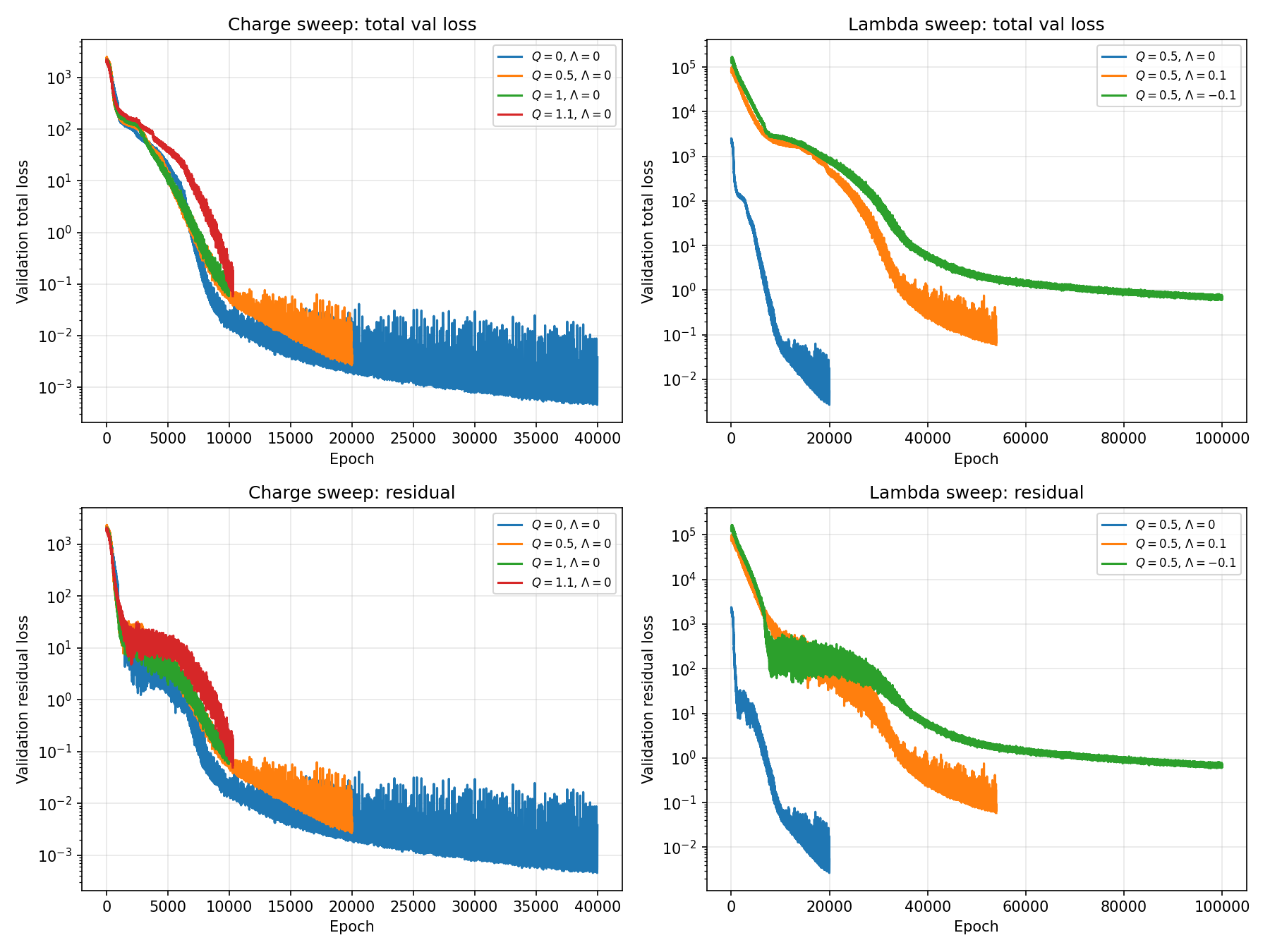}
    \caption{Validation loss histories grouped by charge sweep ($\Lambda=0$, top row)
    and $\Lambda$ sweep ($Q=0.5$, bottom row).}
    \label{fig:loss}
\end{figure}

Figure~\ref{fig:loss} shows validation total, residual, and boundary losses from
TensorBoard for the charge sweep ($\Lambda=0$) and the $\Lambda$ sweep ($Q=0.5$).
Residual loss decreases monotonically after an initial transient; boundary loss
dominates early training and is jointly minimized with the residual. Configurations
with larger $|\Lambda|$ or near-extremal charge require more epochs before early
stopping triggers.

\subsection{Robustness across random seeds}

\begin{figure}[htbp]
    \centering
    \includegraphics[width=0.85\linewidth]{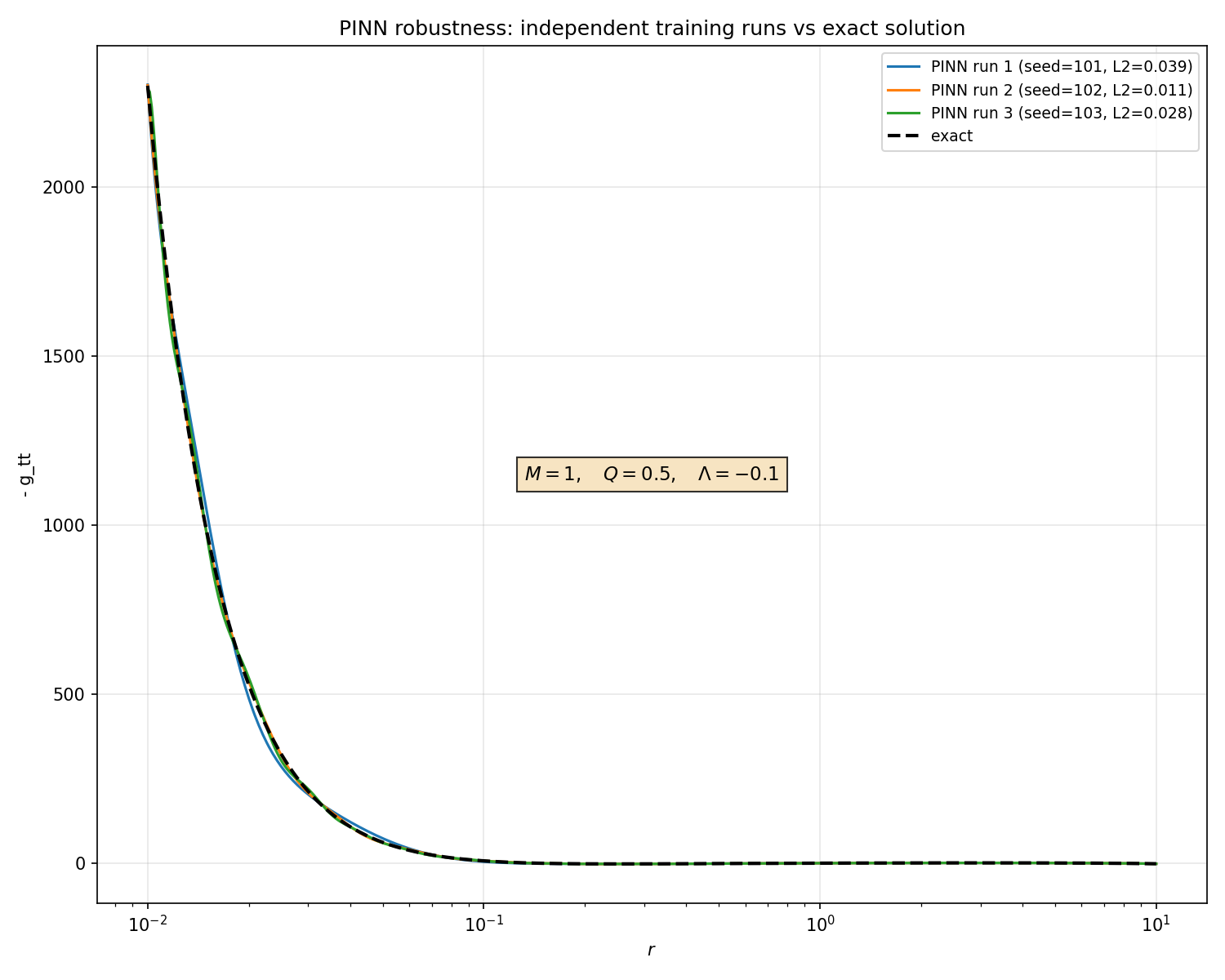}
    \caption{Three independent PINN trainings versus the analytic reference for
    $M=1$, $Q=0.5$, $\Lambda=0.1$. Legend entries report seed and
    $\varepsilon_{\mathrm{rel}}$.}
    \label{fig:robustness}
\end{figure}

For $Q=0.5$, $\Lambda=0.1$, three independent trainings (seeds $101$--$103$) yield
the errors in Table~\ref{tab:robustness}. The spread (standard deviation
$1.41\times 10^{-2}$) is modest relative to the mean ($2.58\times 10^{-2}$), and
all runs remain below the $6\%$ accuracy threshold. Figure~\ref{fig:robustness}
shows that each run closely follows the analytic curve on the full domain.

\begin{table}[htbp]
    \centering
    \caption{Robustness study: $M=1$, $Q=0.5$, $\Lambda=0.1$, three fixed seeds.}
    \label{tab:robustness}
    \begin{tabular}{@{}ccccc@{}}
        \toprule
        Run & Seed & $\varepsilon_{\mathrm{rel}}$ & Val.\ loss & Epoch \\
        \midrule
        1 & 101 & 0.0389 & $5.59\times 10^{-2}$ & 99\,146 \\
        2 & 102 & 0.0108 & $5.78\times 10^{-2}$ & 99\,500 \\
        3 & 103 & 0.0276 & $1.99\times 10^{-2}$ & 90\,671 \\
        \midrule
        \multicolumn{2}{l}{Mean / Std} &
        $0.0258$ / $0.0141$ & --- & --- \\
        \bottomrule
    \end{tabular}
\end{table}

\section{Discussion}
\label{sec:discussion}

Our reduced ODE formulation trades full Ricci-tensor constraints for a single
radial equation, lowering implementation complexity while retaining the PINN
workflow: mesh-free collocation, hard/soft boundary encoding, and derivative
accuracy from autograd. Relative errors below $6\%$ across sweeps support the
approach for exterior metrics with charge and cosmological constant.

The principal architectural distinction from Ref.~\cite{Li2023} is the unified
network of Sec.~\ref{sec:unified} together with the minimal ansatz of
Sec.~\ref{sec:ansatz}. DPINNs partition the domain to ease local
optimization but introduce interface discontinuities in both the error budget and,
potentially, the reconstructed metric (Fig.~\ref{fig:li2023}). Their radial
truncation at $r=10$ further reduces the problem difficulty by omitting the inner
$1/r$ singularity structure of Eq.~\eqref{eq:reference}; our sweeps demonstrate
that a single network can nevertheless fit this harder domain when the input is
compressed through $\log r$ and collocation is log-spaced. A single network with
shared weights enforces global smoothness by design; our metric overlays
(Fig.~\ref{fig:comparison}) show no interior kinks even for charged and
$\Lambda$-deformed exteriors where DPINN error profiles exhibit sharp steps. Training
one approximator over three decades in $r$---from $10^{-2}M$ through the
outer boundary---is substantially more demanding than fitting $r\in(10,300)$ in
patches, but it is also more informative: success here implies the PINN has learned
the global radial structure rather than a mild outer profile alone. Additive skip
connections in each hidden block are a deliberate architectural choice to aid this
convergence: they shorten gradient paths through the six-layer $\tanh$ stack and
help the optimizer reconcile the inner $1/r$ regime with the outer boundary
constraint without resorting to subdomain decomposition. This comes at
the cost of longer training and careful early stopping on validation loss, but
removes the need to tune subdomain count, overlap, interface penalties, or an
inner radial cutoff.

A separate modeling choice concerns how much physics is injected into the network
\emph{before} optimization. Ref.~\cite{Li2023} embed the weak-field Schwarzschild
tail directly in $g_{00}$ through Eq.~\eqref{eq:li-ansatz}, so the network
primarily learns a correction to a nearly correct metric. We instead impose only
leading asymptotic behavior at $r_{\max}$ via Eq.~\eqref{eq:bc} and rely on the
ODE residual to determine the interior. This is a stronger test of the PINN
formalism---especially across charged and $\Lambda$-deformed exteriors where a
Schwarzschild-specific output template would be inappropriate---and it keeps the
learned field closer to a ``pure'' physics-informed solution.

Compared to full metric reconstruction~\cite{Li2023,Luna2022,Patel2024}, we do not yet enforce all
components of $g_{\mu\nu}$ or the full Einstein system; extending to $g_{rr}$ and
Ricci residuals is a natural next step. The $\Lambda=-0.1$ case exhibits the
largest validation loss ($0.59$) yet still achieves $\varepsilon_{\mathrm{rel}}
\approx 5.6\%$, suggesting that loss magnitude and field error are not always
monotonically aligned---monitoring both, as we do with TensorBoard and
Eq.~\eqref{eq:l2}, is advisable.

Seed sensitivity is visible (run~2 reaches $1.08\%$ while run~1 reaches $3.89\%$)
but not catastrophic; fixed-seed reproducibility is useful for regression testing
and manuscript evidence.

\section{Conclusion}
\label{sec:conclusion}

We trained a \emph{unified} physics-informed neural network to recover $-g_{tt}$ for
static black-hole exteriors with variable charge and cosmological constant at $M=1$,
from $r_{\min}=10^{-2}M$ through the outer boundary. This domain includes the
steep inner curvature regime that Ref.~\cite{Li2023} exclude by starting at
$r=10$, while our unified architecture avoids the subdomain discontinuities of
their DPINN formulation. By using the minimal ansatz $A=u/r^2$ and asymptotic
constraints only in the loss---rather than hard-coding metric terms such as
$2M/r-1$ into the network output---the same template succeeds across charge and
$\Lambda$ sweeps without problem-specific architectural redesign. Six primary
configurations and three robustness repeats demonstrate consistent, smooth
agreement with analytic references, with documented loss histories and relative
$\mathcal{L}_2$ metrics. The method is mesh-free, differentiable end-to-end, and
straightforward to extend toward full Einstein PINNs and time-dependent numerical
relativity prototypes.

\end{document}